# CMOS pixel sensors optimized for large ionizing dynamic

W.Ren,[a,b] J.Baudot,[a,1] L.Federici,[c] C.Finck,[a] C.Hu-Guo,[a] M.Kachel,[a] C.-A.Reidel,[d] C.Schui,[d] R.Sefri,[a] E.Spiriti,[c] U.Weber[d] and Y.Zhao[a]

[a] *Institut Pluridisciplinaire Hubert CURIEN (IPHC),*
   *23 rue du Loess, 67307 Strasbourg, France*

[b] *PLAC, Key Laboratory of Quark & Lepton Physics (MOE), Central China Normal University,*
   *152 Luoyu Road, 430079 Wuhan, China*

[c] *Istituto Nazionale di Fisica Nucleare, Laboratori Nazionali di Frascati,*
   *I-00040, Italy*

[d] *GSI Helmholtzzentrum für Schwerionenforschung GmbH,*
   *Planckstrasse 1, 64291Darmstadt, Germany*
   *E-mail*: jerome.baudot@iphc.cnrs.fr

ABSTRACT: Monolithic active pixel sensors (MAPS) are now well established as a technology for tracking charged particles, especially when low material budget is desirable. For such applications, sensors focus on spatial resolution and pixels with digital output or modest charge measurement ability are well suited. Within the European Union STRONG-2020 project, which focuses on experiments using hadrons, the TIIMM (Tracking and Ions Identifications with Minimal Material budget) joint research activity intends to expand granular MAPS capacity to energy-loss (ΔE) measurement for ion species identification. The TIIMM prototypes are developed in the Tower Jazz 180 nm CMOS image sensor (CIS) process. The Time-Over-Threshold (ToT) method is applied to the sensor for the energy-loss measurement. The main design details and the preliminary test results from laboratory measurements of the initial TIIMM prototype are presented in this work.

KEYWORDS: Monolithic active pixel sensors (MAPS); Energy loss measurement; Time-Over-Threshold (ToT).

---

[1] Corresponding author.

**Contents**

**1. Introduction**

The TIIMM (Tracking and Ions Identifications with Minimal Material budget) joint research activity of the European Union STRONG-2020 project aims to create a new class of instrument combining precision tracking and energy loss measurement with minimal material budget. This development is in particular motivated by the need to distinguish different ion species when measuring the heavy ions fragmentation cross-sections [1]. In high-energy physics, Monolithic Active Pixel Sensors (MAPS) are used for charged particle tracking [2][3]. More recently they have also been used in ions fragmentation cross-section measurements for hadron therapy treatment improvement [4]. The TIIMM goal is to complement the excellent tracking capabilities in current MAPS with the measurement of the energy lost in the sensor to identify the crossing ion species [5].

TIIMM sensors target signals from minimum ionization particles up to heavily ionization ions, such as carbon at few 100s MeV/u. Such MAPS will combine tracking and identification capabilities. The challenge lies in the implementation of a dynamic range of the order of $1:10^3$ within a pixel pitch of about 40 µm. TIIMM exploits recent advances in CMOS pixel sensors, where the sensitive layer is fully depleted for more efficient charge collection. Two prototypes have been developed to investigate the feasibility of the time-over-threshold (ToT) method for large dynamic energy-loss measurement.

This contribution details the analog pixel design and initial test results.

**2. The TIIMM prototypes**

The Tower Jazz 180 nm CMOS image sensor (CIS) process has been chosen for the TIIMM project. A first sensor prototype, TIIMM-0, has been fabricated in 2020, while a second sensor, TIIMM-1, is being submitted to the foundry at the end of 2021. Both prototypes have common features: 2.2×1.5 mm² total area, 32×16 pixel matrix, and a similar pixel architecture described in the next subsection.

A standard I2C interface is applied for the slow control of the chips, including the matrix configuration. Thus, it is possible to inject test pulses to the pixel analog part or the in-pixel digital logic. A simple serial readout is implemented in the sensors.

10 DACs are implemented for biasing the pixel front-end circuits. The analog outputs of the last column are connected to an output pad through an analog buffer, allowing the analog output investigation.

**2.1 Pixel design and operation**

Each pixel is composed of a charge collection diode, a Charge Sensitive Amplifier (CSA), a comparator, and a digital logic providing the 6-bit ToT measurement. The comprehensive schematics for the pixel cells are shown for TIIMM0 and TIIMM1 in **Figure 1**. The collection diode is an octagonal-shaped n-well with 1.3 µm diameter, in contact with the highly resistive p-type epitaxial layer. The diode is surrounded by p-wells with a spacing of 3.5 µm. The CSA is composed of an amplifier and the Krummenacher feedback circuit [6]. When the pixel is fired, the charges collected by the diode generate a signal amplified by the CSA. If the output signal of the CSA exceeds the threshold set to the comparator, the output of the comparator will flip. Then the pulse width of the comparator is digitized over 6 bits by the Time-Over-Threshold (ToT) [7] in-pixel digital logic.

The early TIMM0 prototype was designed to validate the pixel architecture. The pixel area is exactly 40×40 µm$^2$. It includes in each pixel a 4-bit local DAC for tuning the threshold.

For the TIIMM1 sensor, the feedback capacitor $C_f$ was increased from 1fF to 5fF to enlarge the dynamic range. The trimming DAC presented in the TIIMM0 pixel was replaced by AC-coupling the CSA output to the comparator to mitigate offset problems. This solution also decreases the power consumption as well as gains extra area to optimize the transistor size improving pulse width fluctuation. The pixel size increases only slightly to 41.2×40 µm$^2$.

Post layout simulation results of both pixel cells are discussed below. From **Figure 2**, we observe that the dynamic for the pulse duration is much larger than the pulse amplitude and the linear range in TIIMM1 can reach 250 ke$^-$ against 110 ke$^-$ for TIIMM0. The AC-coupling also drastically decreases the baseline spread from 9.27 mV to 1.11 µV. These improvements are obtained with a modest ENC increase from 42 e$^-$ (TIIMM0) to 78 e$^-$ (TIIMM1). Finally, the predicted relative fluctuation of the pulse width does not exceed 10% over the whole dynamic range for TIIMM1 while it reaches up to 30% for TIIMM0.

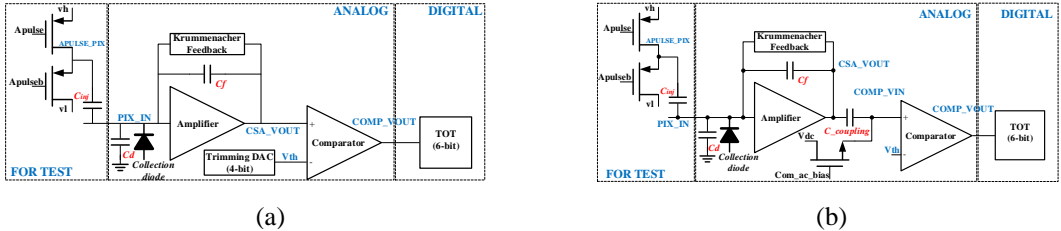

(a)                          (b)

**Figure 1.** (a) Pixel structure of TIIMM0. (b) Pixel structure of TIIMM1.

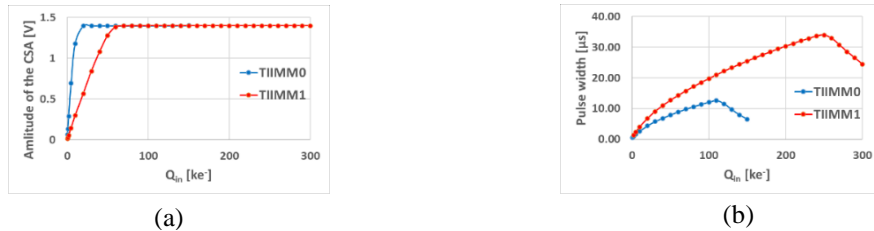

(a)                          (b)

**Figure 2.** Results from post-layout simulations: (a) CSA output (b) pulse width of the comparator vs the input charge.

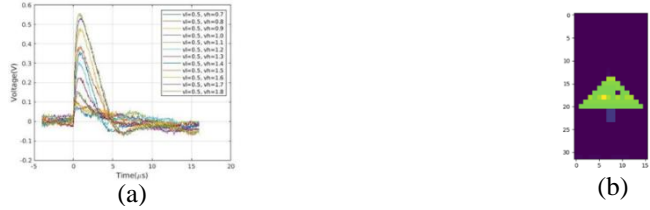

(a)                             (b)

**Figure 3.** (a) The analog ouput of the CSA under different injections. (b) Image with digital readout test. This image is created with digital test signals injected into the pixels.

### 2.2 Preliminary test results with TIIMM

TIIMM0 sensor was fabricated in 2020 and samples were wire-bonded on a custom-designed PCB and characterized in the laboratory to study the chip functionalities.

As shown in **Figure 1**(a), an injection capacitor $C_{inj}$ is implemented in each pixel. The pulse width of the CSA analog output signals increases with large injections as depicted in **Figure 3**(a). Due to the small $C_{inj}$, the maximum injection is around 4 ke$^-$. For performance under larger injections, tests under the radiation source and the beams will be carried out.

**Figure 3**(b) shows the image with the digital readout test. A digital pulse signal with different pulse width is injected into each pixel directly as the input of the ToT module. This image shows the difference of the pulse width for each pixel which means that the ToT module works properly.

### 3. Conclusion

The ToT concept to measure energy loss has been validated with the TIIMM0 prototype. The next sensor benefits from an optimized design for larger dynamic and smaller pulse width relative fluctuation. TIIMM1 with depleted sensitive layer will be available in 2022 for performance tests with various ion beams.

### Acknowledgments

This project has received funding from the European Union's Horizon 2020 research and innovation program under grant agreement No. 824093. The support provided by China Scholarship Council (CSC) No. 201906770018 is acknowledged.